\documentclass[revtex4]{emulateapj}

\usepackage{amssymb}
\usepackage{amsmath}

\tightenlines

\begin{document}

\title{Impulsive spot heating and thermal explosion of interstellar grains revisited}

\author{A. V. Ivlev$^1$, T. B. R\"ocker$^1$, A. Vasyunin$^{1,2}$, and P. Caselli$^1$}
\email[e-mail:~]{ivlev@mpe.mpg.de} \affiliation{$^1$Max-Planck-Institut f\"ur Extraterrestrische Physik, 85741 Garching,
Germany\\ $^2$Ural Federal University, Ekaterinburg, Russia}

\begin{abstract}
The problem of impulsive heating of dust grains in cold, dense interstellar clouds is revisited theoretically, with the aim
to better understand leading mechanisms of the explosive desorption of icy mantles. It is rigorously shown that if the
heating of a reactive medium occurs within a sufficiently localized spot (e.g., heating of mantles by cosmic rays), then the
subsequent thermal evolution is characterized by a single dimensionless number $\lambda$. This number identifies a
bifurcation between two distinct regimes: When $\lambda$ exceeds a critical value (threshold), the heat equation exhibits
the explosive solution, i.e., the thermal (chemical) explosion is triggered. Otherwise, thermal diffusion causes the
deposited heat to spread over the entire grain -- this regime is commonly known as the whole-grain heating. The theory
allows us to find a critical combination of the physical parameters that govern the explosion of icy mantles due to
impulsive spot heating. In particular, the calculations suggest that heavy cosmic ray species (e.g., iron ions) colliding
with dust are able to trigger the explosion. Based on the recently calculated local cosmic-ray spectra, the expected rate of
the explosive desorption is estimated. The efficiency of the desorption, which affects all solid species independent of
their binding energy, is shown to be comparable with other cosmic-ray desorption mechanisms typically considered in the
literature. Also, the theory allows us to estimate maximum abundances of reactive species that may be stored in the mantles,
which provides important constraints on available astrochemical models.
\end{abstract}


\maketitle

\section{Introduction}

The earliest stages of star formation occur in cold ($T\sim10$~K), dense ($n(H)\gtrsim10^4$~cm$^{-3}$), and dark
($A_V\gtrsim10$~mag) molecular cloud cores~\citep[e.g.,][]{Myers1987}. Under such physical conditions, rapid freeze-out of
molecular species from the gas phase on interstellar grains should occur on a timescale of $\sim10^9/n(H)$~yrs. However,
while infrared observations confirm the existence of thick icy mantles on interstellar grains~\citep[e.g.,][]{Gibb_ea2004},
molecular species are also observed in the dark cold gas~\citep[][]{Tafalla_ea2002,Caselli_ea2002,
Tafalla_ea2004,Caselli2012a}. As the lifetime of cold molecular cores is at least $\sim10^6$~yr \citep[][]{Brunken_ea2014},
a non-thermal desorption mechanism is required to maintain the observed gas-phase abundances of species. The recent
discovery of complex organic molecules \citep[][]{Oeberg_ea2010, Bacmann_ea2012, Cernicharo_ea2012} and deuterated methanol
\citep{Bizzocchi2014} in the cold gas is further evidence for non-thermal processing and evaporation of cold icy
mantles~\citep[][]{VasyuninHerbst2013a}.

Interactions of interstellar grains with cosmic ray (CR) particles, X-ray and UV photons, and even their mutual collisions
cause the grain heating and hence stimulate sublimation of ice \citep{dHendecourt1982,Leger1985, Hartquist1990,Schutte1991,
Hasegawa1993,Shalabiea1994, Shen2004,Bringa2004, Cuppen2006,Herbst2006, Roberts2007}. Depending on the mechanism of energy
deposition, the heated region may be localized or it may extend over the entire grain -- these two scenarios are usually
referred to as ``spot heating'' and ``whole-grain heating'', respectively \citep{Leger1985,Schutte1991,Shen2004,Bringa2004}.
Also, some exothermic reactions occurring on the grain surface (e.g., the formation of molecular hydrogen) may result in the
local heating and lead to the chemical desorption of weakly bound species \citep{Duley1993,Garrod2007,
Cecchi-Pestellini2012, Rawlings2013}.

One can identify two distinct regimes of desorption occurring in response to the impulsive grain heating: The classical
thermal evaporation, and the so-called ``explosive desorption'' triggered by the exothermic chemical reaction(s) between
free radicals frozen in the bulk of ice \citep{dHendecourt1982,Leger1985,Schutte1991,Shalabiea1994}. The essential
difference between the two regimes is that the evaporation of the ice mantle (typically limited to the most volatile
species) is accompanied by the grain cooling, whereas the chemical reactions (activated by the deposited energy) can lead to
the runaway temperature growth. As a result, the explosive desorption may cause the ejection of the entire mantle off the
grain surface.

Since the 1980's, there have been various mechanisms proposed to trigger the thermal explosion of icy mantles. In
particular, these include inelastic collisions between the grains, when a certain fraction of their kinetic energy is
converted into heat in the mantle \citep{dHendecourt1982,Schutte1991,Shalabiea1994}, and the impact of energetic particles,
such as CR and X rays \citep{Leger1985,Shen2004}. The analysis, however, has been {\it almost completely} focused on the
whole-grain-heating scenario, neglecting the initial thermal spikes emerging in a grain (e.g., along the CR paths). To the
best of our knowledge, the possibility of thermal explosion due to CR spot heating was only discussed by \cite{Leger1985},
who concluded that such process is not feasible.\footnote{We note that non-explosive desorption due to spot heating has been
extensively studied \citep[e.g.,][]{Leger1985,Shen2004,Bringa2004}.}

In this article we revisit the problem of spot heating of interstellar grains. We introduce a concept of the {\it localized}
ignition spot and show that the evolution of the initial kinetic energy deposited in a reactive medium in this case is
uniquely described by a {\it single} dimensionless number $\lambda$. This concept allows us to calculate a critical value of
$\lambda$ above which the thermal explosion is triggered and, hence, to find a critical combination of the physical
parameters that govern the explosion of icy mantles due to spot heating. We show that the energy deposited by iron CR are
sufficient to cause such explosions. Furthermore, we demonstrate that the chemical explosion due to whole-grain heating is
inhibited by efficient sublimation cooling. Based on the recent calculations of local CR energy spectra, we obtain the
minimum expected rate of mantle disruption due to impacts of iron CR. Finally, the presented theory allows us to estimate
the maximum abundances of reactive species that may be stored in the mantles and, thus, to impose important constraints on
available astrochemical models.

\section{Theory}
\label{theory}

Consider the situation when a certain amount of kinetic energy is ``instantaneously'' deposited in a reactive medium. It is
intuitive to expect that the exact form of the initial energy distribution must be unimportant for its subsequent evolution,
provided this distribution is sufficiently localized {\it and} the energy is rapidly thermalized. Mathematically, the
possibility of the thermal explosion in this case can be investigated by assuming the initial temperature distribution in
the form of the delta function. Limits of applicability for such approximation of the ignition spot are determined from the
numerical analysis, as discussed below.

Let us consider the cases where the initial energy is concentrated on a plane, along an axis, or in a point. For such
ignition spots, the problem is characterized by the symmetry indices $D=1,2,$ and 3, respectively, and the heat equation
describing the temperature distribution $T(r,t)$ in a reactive medium has the following form \citep{LandauFluid}:
\begin{equation}
\rho c\frac{\partial T}{\partial t}= Q_{\rm r}e^{-E_{\rm a}/k_{\rm B}T}+\kappa\left(\frac{\partial^2T}{\partial r^2}
+\frac{D-1}r \frac{\partial T}{\partial r}\right),\label{heat_eq}
\end{equation}
with the initial condition
\begin{equation}
T(r,0)=\frac{q_D}{\rho c}\delta_D(r).\label{IC}
\end{equation}
Here, $\delta_D(r)$ is the delta function in $D$ dimensions, $q_D$ is the initial energy density in the ignition spot,
$\kappa$, $\rho$, and $c$ are, respectively, the thermal conductivity, mass density, and specific heat of the medium
(treated as incompressible, so $c$ should be taken at constant pressure), $Q_{\rm r}$ is the heat of reaction per unit
volume and time, and $E_{\rm a}$ is the relevant activation energy in the Arrhenius factor (definition of $Q_{\rm r}$ and
proper choice of $E_{\rm a}$ are discussed in Sec.~\ref{properties}). We first assume the properties of the medium to be
independent of the temperature -- the cases when $c$ or/and $\kappa$ are functions of $T$ are considered later.

\begin{figure}\centering
\includegraphics[width=.9\columnwidth,clip=]{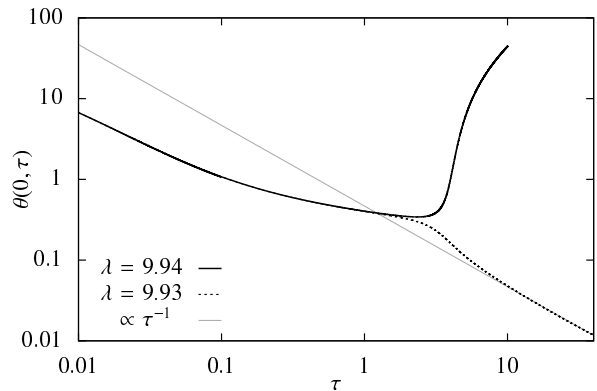}
\caption{Evolution of the dimensionless temperature at the center of the ignition spot, $\theta(0,\tau)$. Shown are
numerical solutions of Eqs.~(\ref{heat_eq_norm}) and (\ref{IC_norm}) for $D=2$, obtained for marginally under-critical
(dashed line) and over-critical (solid line) values of $\lambda$. The under-critical curve tends asymptotically to the
solution for $\lambda=0$, $\theta(0,\tau)\propto\tau^{-D/2}$ (thin solid line), the over-critical curve approaches
$\theta(0,\tau)\propto\tau$ after the explosive growth. For $D=1$ and 3 a similar behavior is observed.}
\label{fig1}
\end{figure}

\begin{figure}\centering
\includegraphics[width=.9\columnwidth,clip=]{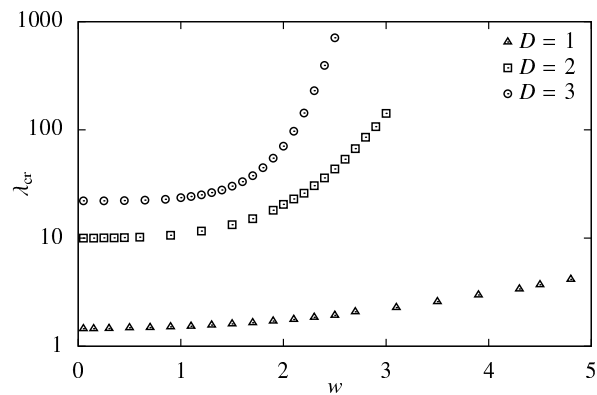}
\caption{Dependence of the numerically calculated explosion threshold, $\lambda_{\rm cr}$, on the normalized size of the
ignition spot, $w$. The thresholds, plotted for $D=1,2,$ and 3, are nearly constant for $w\lesssim1$, and rapidly increase
at larger $w$ (see Appendix~\ref{App_relation} for details). The numerical accuracy for $\lambda_{\rm cr}$ is better than
$\pm5\%$.}
\label{fig2}
\end{figure}

\begin{figure}\centering
\includegraphics[width=.9\columnwidth,clip=]{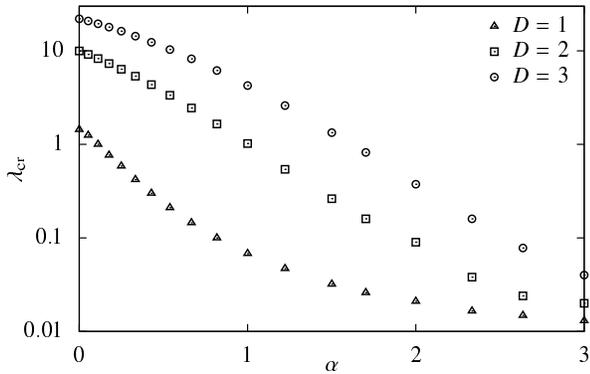}
\caption{Effect of the temperature-dependent specific heat on the explosion threshold. The critical numbers
$\lambda_{\rm cr}$, numerically calculated for $D=1,2,$ and 3, are plotted versus the exponent $\alpha$ determining the
temperature dependence $c\propto T^{\alpha}$. The numerical accuracy for $\lambda_{\rm cr}$ is better than $\pm5\%$.}
\label{fig3}
\end{figure}

For the analysis of Eqs.~(\ref{heat_eq}) and (\ref{IC}) we normalize the temperature by the activation energy,
$\theta=k_{\rm B}T/E_{\rm a}$. For the dimensionless distance $\xi=r/r_*$ we choose the scale $r_*$ which provides unity
normalization of Eq.~(\ref{IC}), while the dimensionless time $\tau=t/t_*$ is determined by the timescale $t_*$ of thermal
diffusion at the distance $r_*$. This yields
\begin{equation}\label{scales}
r_*=\left(\frac{q_D}{\rho cE_{\rm a}}\right)^{1/D},\quad t_*=\frac{\rho c}{\kappa}r_*^2,
\end{equation}
so the heat equation is reduced to
\begin{equation}\label{heat_eq_norm}
\frac{\partial \theta}{\partial \tau}= \lambda e^{-1/\theta}+\frac{\partial^2\theta}{\partial \xi^2}
+\frac{D-1}{\xi}\frac{\partial \theta}{\partial \xi},
\end{equation}
and the initial condition -- to
\begin{equation}\label{IC_norm}
\theta(\xi,0)=\delta_D(\xi).
\end{equation}
Thus, in the dimensionless form the problem is characterized by a {\it single} number,
\begin{equation}\label{lambda}
\lambda=\frac{Q_{\rm r}}{\kappa E_{\rm a}}\left(\frac{q_D}{\rho cE_{\rm a}}\right)^{2/D}.
\end{equation}
The role of $\lambda$ is similar to that of the Frank-Kamenetskii number $\lambda_{\rm FK}$ [see Eq.~(\ref{lambda_FK})]
which governs the thermal stability of a steady state \citep{FrankKamenetskii,LandauFluid}. The relation between $\lambda$
and $\lambda_{\rm FK}$ is discussed in Appendix~\ref{App_relation}.

The thermal explosion is triggered when $\lambda$ exceeds a certain critical value $\lambda_{\rm cr}$ -- the explosion
threshold. From the numerical solution of Eqs.~(\ref{heat_eq_norm}) and (\ref{IC_norm}) we obtain the following thresholds:
\begin{eqnarray}
  D=1: &\quad& \lambda_{\rm cr}=1.45;\nonumber\\
  D=2: &\quad& \lambda_{\rm cr}=9.94;\nonumber\\
  D=3: &\quad& \lambda_{\rm cr}=22.1.\nonumber
\end{eqnarray}
The bifurcation between the decaying and explosive evolutions is illustrated for $D=2$ in Fig.~\ref{fig1}, where the
temperature at the center of the ignition spot, $\theta(0,\tau)$, is plotted. For $\lambda<\lambda_{\rm cr}$ the integral
effect of thermal diffusion is stronger than that of reaction heating, so the asymptotic temperature decay is described by
the fundamental solution of the heat equation in free space \citep{LandauFluid}, which yields
$\theta(0,\tau)\propto\tau^{-D/2}$. When $\lambda>\lambda_{\rm cr}$, thermal diffusion becomes asymptotically negligible and
the temperature approaches linear growth, since the Arrhenius term in Eq.~(\ref{heat_eq_norm}) tends to a constant
($\lambda$) for large $\theta$. In Fig.~\ref{fig1}, the bifurcation occurs at $\tau\sim3$ (while for $D=1$ and 3 it is at
$\tau\sim10$ and $\sim1$, respectively). We conclude that the explosion develops within the physical time of a few $t_*$.

For the numerical solution, we approximate the initial energy distribution by a rectangular function with width $w$ (the
delta function formally corresponds to the limit $w\to0$). The obtained dependence $\lambda_{\rm cr}(w)$ is plotted in
Fig.~\ref{fig2}, showing that the explosion thresholds remain practically constant for $w\lesssim1$. Thus, the problem does
not (practically) depend on the physical size of the ignition spot as long as it is smaller than $\sim r_*$, i.e., the
initial energy distribution for such {\it localized} spots is well represented by the delta function.

Once the explosion is triggered, the hot reactive zone starts expanding away from the ignition spot. As discussed in
Appendix~\ref{App_front}, the flame front propagates with a constant speed $U$ determined by Eq.~(\ref{U}). In
Sec.~\ref{imp1} we demonstrate that the magnitude of $U$ is much smaller that the typical sound speed in solids.

The above results can be generalized for the case where properties of the medium depend on the temperature. In
Appendix~\ref{App_T_dependence} we show that for a power-law temperature dependence of the specific heat, $c(T)\propto
T^{\alpha}$ (typical for solids, see Sec.~\ref{properties}), the explosion threshold rapidly decreases with the exponent
$\alpha$. Figure~\ref{fig3}, illustrating the case $D=2$, demonstrates that for the linear temperature dependence the value
of $\lambda_{\rm cr}$ decreases by one order of magnitude, and for the quadratic -- by two. Note that the number $\lambda$
as well as the front speed $U$ in this case are given by Eqs.~(\ref{lambda1}) and (\ref{U1}). We also demonstrate that the
temperature dependence of the thermal diffusivity $\chi=\kappa/\rho c$ has relatively weak effect on the results.

\section{Implication for interstellar dust grains}
\label{implications}

\begin{figure*}\centering
\includegraphics[width=.8\textwidth,clip=]{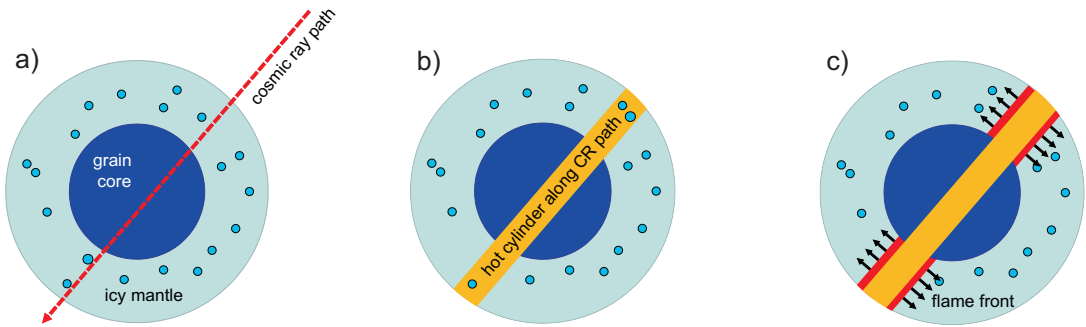}
\caption{Schematic representation of the chemical explosion of icy mantle due to cosmic ray impact. (a) A refractory core of
a grain is covered by a thick icy mantle, where some reactive species (radicals, blue dots) are stored in the bulk. Each
collision of a CR particle with a grain is accompanied by loss of energy, deposited along the CR path. (b) This creates a
hot, narrow cylindrical region whose subsequent evolution is governed by the dimensionless number $\lambda$, given by
Eq.~(\ref{lambda}): If $\lambda$ is below a certain critical value, the deposited energy is simply redistributed over the
grain's volume (the whole-grain-heating scenario). Otherwise, the thermal explosion is triggered (c) -- runaway exothermic
reactions generate a cylindrical flame front in the mantle, leading to its disruption.} \label{fig4}
\end{figure*}

In this section, the theory presented in Sec.~\ref{theory} is applied to interstellar dust grains, to obtain conditions when
impulsive heating by energetic particles is expected to cause the thermal explosion of icy mantles.

The impulsive heating by CR particles, sketched in Fig.~\ref{fig4}, has the axial symmetry and is described by the solution
for $D=2$. The initial energy density $q_2$, which enters the dimensionless number $\lambda$ in this case, is equal to the
stopping power of a CR particle. The stopping power depends on the particle kinetic energy $\varepsilon$ (per nucleon) and
exhibits a broad maximum at $\varepsilon=\varepsilon^{\rm max}$ \citep{ZieglerBook}: for protons, $\varepsilon_{\rm H}^{\rm
max}\sim0.1$~MeV and $q_2(\varepsilon_{\rm H}^{\rm max}) \sim 10^{-10}$~J~cm$^{-1}$; for iron ions, $\varepsilon_{\rm
Fe}^{\rm max}\sim1$~MeV/nucleon and $q_2(\varepsilon_{\rm Fe}^{\rm max}) \sim10^{-8}$~J~cm$^{-1}$. The heating by X rays is
better described by the spherically-symmetric solution, $D=3$ (see discussion in Sec.~\ref{imp1}).

\subsection{Properties of icy mantles}
\label{properties}

Let us summarize typical physical properties of mantles which determine the magnitude of $\lambda$.

For many amorphous solids (including ice) the specific heat $c$ increases approximately as $\propto T^2$ at lower
temperatures, with $10^{-2}$~J~cm$^{-3}$K$^{-1}$~$\lesssim\rho c\lesssim0.3$~J~cm$^{-3}$K$^{-1}$ for $10$~K~$\leq T\leq
50$~K; the growth becomes slower at higher temperatures, $\rho c\sim3$~J~cm$^{-3}$K$^{-1}$ at $T\sim10^3$~K
\citep{Zeller1971,Leger1985}. We employ this generic dependence for the estimates below. For the thermal conductivity
$\kappa=\rho c\chi$ we use the diffusivity $\chi\sim10^{-2}$~cm$^2$s$^{-1}$ \citep{dHendecourt1982,Leger1985,Schutte1991};
the latter is approximately constant for many amorphous solids at $T\gtrsim30$~K \citep{Zeller1971}. Note that $\chi$ may
decrease with $T$ for amorphous water ice \citep{Andersson2002}, but this should only have a minor effect on results (see
Appendix~\ref{App_T_dependence}).

For the sake of clarity we suppose that among the variety of reactive species (radicals) stored in the mantle, there is a
pair (A and B) whose exothermic reaction dominates the heat release (the approach can be straightforwardly generalized to
multiple reactions). The heat rate is then given by \citep{Leger1985} $Q_{\rm r}\simeq E_{\rm r}\varphi_{\rm A}\varphi_{\rm
B}N\nu$, where $E_{\rm r}\sim3$~eV is the typical energy release per reaction, $\varphi_{\rm A,B}=N_{\rm A,B}/N$ are the
fractional abundances of the species, $\nu\simeq2\times10^{12}$~s$^{-1}$ is their characteristic vibration frequency, and
$N\simeq3\times 10^{22}$~cm$^{-3}$ is the total number density of molecules in ice \citep{Leger1985,Schutte1991}. Thus, to
estimate the magnitude of $Q_{\rm r}$ we need to know the abundance of reactive species. Let us elaborate on this point.

Direct infrared observations of interstellar ices can only supply us with the abundances of major ice constituents which are
in general not reactive under cold ISM conditions, with the exception of CO ice. As such, we have to rely on astrochemical
modeling when estimating abundances of reactive species in a typical interstellar ice. Early astrochemical models did not
have a distinction between reactive surface and more inert bulk of a thick icy mantle~\citep[][]{Hasegawa_ea1992,
Hasegawa1993}. Therefore, in these models all species adsorbed on a grain surface participate in efficient ``surface''
chemistry and the resulting fraction of radicals stored in the mantle is very low. However, in a number of more recent
studies, several important effects were recognized that favor larger amounts of radicals to be stored in the interstellar
ice. First, icy mantles in dark clouds are likely to be thick and consisting of several hundreds of monolayers \citep[see,
e.g., Sec.~4.2 of][]{Caselli2012b}. For this reason, reactive species in the inner layers of the ice may be quickly covered
by new accreting species during the ice formation, and become excluded from the rapid surface chemistry. Reactive species
become frozen into water ice and thus survive and accumulate~\citep[e.g.,][]{Taquet2012}. Moreover, it is likely that icy
mantles are exposed to the UV photons even in dark clouds. Photons can penetrate the entire mantle and dissociate stable
molecules in the ice \citep{Cruz-Diaz2014,Chang2014}, thus producing radicals \citep{Garrod2013}. Finally, the amount of
radicals in the ice may be affected by the internal ice structure: theoretical studies show that the porous structure of ice
favors accumulation of the radicals~\citep[][]{Taquet2012}. However, some authors show that interstellar ices are rather
compact than porous~\citep[][]{Garrod2013a}.

To obtain quantitative estimates of the fraction of reactive species stored in the ice in a dark cloud, we simulate the
formation of the icy mantle during the contraction of a diffuse cloud into a dense core using our MONACO code and a simple
evolutionary model presented in \citet{VasyuninHerbst2013b}. Briefly, in the evolutionary model, the temperature linearly
decreases with time from 20~K to 10~K, and the gas density increases from $10^3$~cm$^{-3}$ in the beginning to
$10^5$~cm$^{-3}$ at the end of the contraction. Visual extinction $A_V$ increases self-consistently with the density, from
$A_V=3$ to $A_V\geq10$. The MONACO code has been updated in comparison to \citet{VasyuninHerbst2013b}, and now it includes
chemistry in the bulk, due to ice photoprocessing and intramantle diffusion of species (details will be described in a
future paper).

Mobility of species in the bulk of ice is likely to be significantly lower than on the surface, due to the larger number of
neighboring species that bonded to each other~\citep[][]{Garrod2013}. In our microscopic formalism, this means higher
diffusion energy for a species in the bulk than on the surface. Following \citet{Garrod2013}, we set the diffusion energy of
species in the bulk to be two times the respective surface diffusion energy. The latter, in turn, is usually taken as a
fraction of the sublimation enthalpy \citep[typically, their ratio varies from 0.3 to 0.8, see,
e.g.,][]{Hasegawa_ea1992,Ruffle2000}, here we chose the value of 0.5 in agreement with the best-fit model by
\cite{VasyuninHerbst2013b}. As such, the diffusion energy is $\simeq1150$~K for CO molecules, so CO as well as other
abundant reactive species (with higher diffusion energies) that are produced during ice photoprocessing and entrapment of
accreting material can effectively accumulate in the bulk of ice. Thus, we shall consider the CO diffusion energy as the
relevant activation energy for the Arrhenius factor in Eq.~(\ref{heat_eq}), i.e., $E_{\rm a}/k_{\rm B}=1150$~K.

\begin{figure}\centering
\includegraphics[width=0.97\columnwidth,clip=]{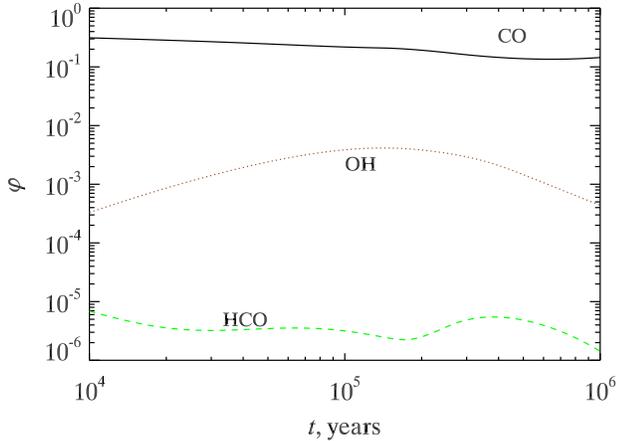}
\caption{Evolution of the fractional abundances of the major reactive species (with respect to the total number of molecules
in the ice). Results are from the numerical modeling with the modified MONACO code.}
\label{fig5}
\end{figure}

In Fig.~\ref{fig5}, the fractional abundances $\varphi$ of the most abundant reactive species in the ice are plotted versus
time. CO is mainly accreted from the gas phase. Some of it undergoes hydrogenation and ultimately converts to methanol and
other saturated species, but significant fraction of CO molecules get buried in the icy mantle in a pristine form. The next
most abundant species is OH, which is mainly produced via dissociation of water by photons and CR protons (according to
\citet{Andersson2008}, only a fraction of the dissociation products recombine back to H$_2$O). Note that the abundance of OH
in our model is about two orders of magnitude lower than in other models of multilayer ice~\citep[e.g.,][]{Taquet2012}.
Presumably, this is due to the fact that we take into account efficient recombination of OH with free H atoms that are
generated in the bulk of ice and perform a random walk before reaching the ice surface. Finally, a certain fraction of HCO
is produced in the bulk, mainly in dissociation of methanol by cosmic ray protons. We see that the abundances of CO and OH
reach the values of $\varphi_{\rm CO}\sim10^{-1}$ and $\varphi_{\rm OH}\sim3\times10^{-3}$ at later stages of the
contraction. We employ these characteristic values for the estimates below.

\subsection{Explosion due to spot heating}
\label{imp1}

In Sec.~\ref{theory} we pointed out that the presented theory can be used as long as the physical size of the ignition spot
does not exceed $\sim r_*$. By substituting typical parameters (listed above, with $\rho c=0.3$~J~cm$^{-3}$K$^{-1}$) in
Eq.~(\ref{scales}) for $D=2$, we obtain $r_*\sim 3\times10^{-6}$~cm for the heating by iron CR. This value is substantially
larger than the diameter of the cylindrical volume where CR deposit their energy \citep[$\lesssim100$~{\AA},][]{Leger1985}
and, at the same time, is smaller than the size of large grains dominating the interstellar dust mass ($\sim10^{-5}$~cm).
Furthermore, the cylindrical explosion develops during the time of the order of $3t_*\sim3\times10^{-9}$~s, which is much
longer than the time during which the deposited CR energy is thermalized \citep[$\lesssim10^{-11}$~s,][]{Leger1985}. Thus,
the theory is indeed applicable to study the reaction of large grains on impulsive heating by heavy CR species (assuming
dust properties that are typically used in astrochemical modeling, see Sec.~\ref{properties}).

Let us estimate the magnitude of $\lambda$ for individual collisions with iron CR. First, we assume a constant $c$ (and
$\kappa$) for icy mantles. By substituting in Eq.~(\ref{lambda}) $q_2\sim10^{-8}$~J~cm$^{-1}$ and $\rho
c=0.3$~J~cm$^{-3}$K$^{-1}$, and setting $E_{\rm r}=3$~eV and $\varphi_{\rm CO}\varphi_{\rm OH}=3\times10^{-4}$ for the
reaction between CO and OH, we obtain $\lambda\sim30$, which exceeds $\lambda_{\rm cr}\simeq10$ for $D=2$. Hence, iron ions
with the energy corresponding to the maximum of the stopping power are able to trigger the explosion,\footnote{The explosion
threshold could also be evaluated by directly comparing the rates of chemical reaction and thermal diffusion in
Eq.~(\ref{heat_eq}). However, as shown in Appendix~\ref{App_relation}, this comparison should be performed at the ``optimum
moment'' (short before the bifurcation in Fig.~\ref{fig1}). Presumably, this latter point was not taken into account by
\cite{Leger1985}, who concluded that the thermal explosion due to spot heating is unlikely.} whereas CR protons with
$q_2\sim10^{-10}$~J~cm$^{-1}$ remain under-critical, since $\lambda\propto q_2$.

Remarkably, when the temperature dependence of the specific heat is taken into account, the resulting ratio
$\lambda/\lambda_{\rm cr}$ becomes even larger, i.e., the explosion condition is relaxed in comparison with the constant-$c$
case: For $c(T)\propto T^{\alpha}$, we employ the results of Appendix~\ref{App_T_dependence} and calculate the enthalpy
scale $H_E$ with $\rho c(E)\sim3$~J~cm$^{-3}$K$^{-1}$ and $1\leq\alpha\leq2$; using the dependence $\lambda_{\rm
cr}(\alpha)$ plotted in Fig.~\ref{fig3}, and substituting $H_E$ in Eq.~(\ref{lambda1}) we obtain $\lambda/\lambda_{\rm cr}$
in the range between $\sim3$ and $\sim30$ for iron CR. Thus, even if some numbers used above for estimating $\lambda$ would
be somewhat less favorable (e.g., if $\varphi_{\rm A}\varphi_{\rm B}\sim3\times10^{-5}$), iron CR should still lead to the
explosion.

The flame front, generated in the mantle by the explosion, propagates with the speed $U$ given by Eqs.~(\ref{U}) or
(\ref{U1}). From this we obtain $U\sim10^4$~cm~s$^{-1}$, which is more than an order of magnitude lower than the typical
sound speed in ice \citep[see, e.g.,][]{Vogt2008}. The crossing time in a grain of the radius $a$ is $\sim a/U\sim10^{-9}$~s
for $a\sim10^{-5}$~cm, so one would expect a practically instant evaporation of the whole mantle. Yet one should keep in
mind that the flame front exerts enormous stress -- the thermal pressure $\sim NT$ substantially exceeds GPa-level, while
the tensile strength of ice is less than one MPa \citep[e.g.,][]{Petrovic2003}. This may lead to mechanical disruption of
the mantle before it is completely evaporated.

One can estimate the rate of mantle disruption due to the thermal explosions, $1/t_{\rm dis}$, which is determined by the
local energy spectrum of iron CR. We assume a constant abundance of iron ions of $\phi_{\rm Fe}\sim10^{-4}$ \citep[relative
to protons, see, e.g.,][]{Leger1985,Shen2004} and employ the local proton spectrum $J_{\rm H}(\varepsilon)$ from
\cite{Padovani2009}, where $\varepsilon$ is the energy per nucleon. The disruption rate is equal to the product of the grain
cross section and the CR flux contributing to the explosion. The minimum value of the latter can be roughly estimated
as\footnote{To obtain the CR flux contributing to the explosion, one should integrate $J_{\rm H}(\varepsilon)$ over the
range of energies (around $\varepsilon_{\rm Fe}^{\rm max}$) where $q_2(\varepsilon)$ exceeds the critical value, i.e., where
$\lambda(q_2)\gtrsim10$. Given the uncertainties in the local spectrum in this range \citep{Padovani2009}, only the lower
bound of the flux can be reasonably estimated.} $\sim4\pi\phi_{\rm Fe}\varepsilon_{\rm Fe}^{\rm max}J_{\rm
H}(\varepsilon_{\rm Fe}^{\rm max})$, where $\varepsilon_{\rm Fe}^{\rm max}\sim1$~MeV/nucleon corresponds to the maximum of
stopping power for iron ions \citep{ZieglerBook}. Even for dense clouds \citep[with the column density of molecular hydrogen
of $\sim3\times10^{22}$~cm$^{-3}$, where the spectrum is strongly attenuated,][]{Padovani2009}, we obtain that the
disruption rate for large grains ($a\sim10^{-5}$~cm) is not lower than
\begin{eqnarray}
\nonumber
1/t_{\rm dis}\sim(2\pi a)^2\phi_{\rm Fe}\varepsilon_{\rm Fe}^{\rm max}J_{\rm H}(\varepsilon_{\rm Fe}^{\rm max})
\sim10^{-6}~{\rm yr}^{-1}.
\end{eqnarray}
Furthermore, supposing the entire mantle evaporated upon disruption, we can also estimate the minimum desorption rate of
molecules into the gas phase. For a mantle with thickness $\Delta a$, the desorption rate of species A is $\sim4\pi
a^2\Delta aN\varphi_{\rm A}/t_{\rm dis}\propto a^5$ (assuming $\Delta a\propto a$). We see that the explosive desorption is
heavily dominated by large grains, with the desorption rate of the order of $3\times10^{-7}$~molecules~grain$^{-1}$s$^{-1}$
for CO molecules. This value is comparable to the desorption rates due to a combination of other mechanisms
\citep{Shen2004,Herbst2006}, such as explosion due to whole-grain heating (see next section for its critical discussion),
and evaporation due to whole-grain and spot heating. We note that the calculations of the desorption rate reported earlier
\citep{Leger1985,Hartquist1990, Hasegawa1993,Shen2004, Bringa2004,Herbst2006} do not take into account attenuation of the
local CR spectrum, which is included in our analysis.

It is noteworthy that we can practically exclude other sorts of energetic particles (e.g., X rays or UV photons) as possible
causes of explosion due to spot heating. To demonstrate this, let us consider X rays as the most energetic species among
such particles: The maximum energy which can be deposited in a grain by an X-ray photon is limited by the condition that the
stopping range of electrons produced by the photon is smaller than the grain size; for $a\sim10^{-5}$~cm we get the upper
energy limit of the order of a few keV \citep[see, e.g.,][]{Leger1985}. Since energetic electrons lose most of the energy at
the end of their paths, the spherically-symmetric solution is more appropriate to describe the problem in this case. For
$D=3$ (and otherwise the same parameters as above), from Eq.~(\ref{lambda}) we obtain that the minimum ignition energy to
satisfy the condition $\lambda>\lambda_{\rm cr}\simeq22$ is $q_3\sim10^5$~eV, which exceeds the maximum deposited energy by
about two orders of magnitude.

Finally, the presented theory allows us to impose important constraints on the fractional abundance of reactive species in
icy mantles, and thus to discriminate between different astrochemical models. In particular, one can estimate the {\it upper
limit} of the abundance of radicals which can be stored in a mantle: For example, some models predict that at later stages
of the cloud evolution, the product of the relative abundances of such radicals may be as high as $\varphi_{\rm
A}\varphi_{\rm B}\sim3\times10^{-3}$ \citep[or even higher, see, e.g.,][]{Schutte1991,Shalabiea1994,Taquet2012,Chang2014}.
Since $\lambda\propto q_2\varphi_{\rm A}\varphi_{\rm B}$, its value for iron CR would then be about two orders of magnitude
larger than $\lambda_{\rm cr}$, so the obtained abundances could already be marginally sufficient to satisfy the explosion
condition for CR protons. However, the latter are $\sim10^{4}$ more abundant than iron CR, so the very possibility of mantle
explosion due to impacts of CR protons would imply unrealistically high disruption rates, of $\sim10^{-3}~{\rm yr}^{-1}$ or
even larger (these exceed the freeze-out rates at typical molecular cloud densities, i.e., icy mantles simply would not have
time to grow). Hence, such high abundances of radicals can be ruled out based on the explosion theory.

\subsection{On the whole-grain heating}
\label{imp2}

If the stopping power of an energetic particle colliding with a grain is too low (under-critical), the deposited energy is
rapidly redistributed over the whole grain. Even though the overall temperature increase in this case could be only a few
tens of degrees, this leads to an exponential amplification of chemical heating in the entire reactive volume of the grain,
with important consequences for surface chemistry and the chemical composition of icy mantles. The thermal stability in this
regime is determined by the global balance between the volume heating and the surface cooling due to thermal radiation and
sublimation \citep{Leger1985,Schutte1991,Cuppen2006}. Therefore, it has been usually argued that there is a critical
temperature of the whole-grain heating, above which the explosion must be triggered
\citep{dHendecourt1982,Leger1985,Schutte1991,Shalabiea1994,Shen2004}. As we pointed out in the introduction, other
mechanisms of the whole-grain heating, e.g., due to inelastic grain-grain collisions, have also been suggested as a possible
cause of the explosion.

Let us consider the global thermal balance for a reactive spherical grain. The steady-state temperature distribution inside
the grain is almost homogeneous, so the heating power $P_{\rm heat}$ is approximately the product of $Q_{\rm r}e^{-E_{\rm
a}/k_{\rm B}T}$ and the reactive (mantle) volume $4\pi a^2\Delta a$. The surface cooling at temperatures above $\simeq25$~K
is dominated by sublimation \citep{Leger1985,Schutte1991}. The resulting cooling power $P_{\rm cool}$ is the product of the
area $4\pi a^2$ and the cooling rate $\Delta H_{\rm sub}(2\pi mk_{\rm B}T)^{-1/2}p_0e^{-\Delta H_{\rm sub}/k_{\rm B}T}$,
where $\Delta H_{\rm sub}$ and $m$ are the sublimation enthalpy and the mass of evaporating molecules, respectively, and
$p_0$ is the pre-factor for the saturated vapor pressure \citep{Leger1983,Leger1985}. By substituting the heat rate $Q_{\rm
r}= E_{\rm r}\varphi_{\rm A}\varphi_{\rm B}N\nu$ for reactive species A and B, we obtain the heating-to-cooling power ratio,
\begin{eqnarray}
\nonumber
\frac{P_{\rm heat}}{P_{\rm cool}}\sim\varphi_{\rm A}\varphi_{\rm B}\frac{E_{\rm r}N\Delta a\nu\sqrt{2\pi mk_{\rm B}T}}{\Delta
H_{\rm sub}p_0}e^{(\Delta H_{\rm sub}-E_{\rm a})/k_{\rm B}T},
\end{eqnarray}
which must exceed unity for the temperature to increase with time. For estimates we set $T=30$~K which ensures that,
irrespective of poorly known emission efficiency of grains, the radiative cooling is negligible
\citep{Leger1985,Schutte1991,Shen2004}. By adopting $E_{\rm a}/k_{\rm B}\simeq\Delta H_{\rm sub}/k_{\rm B}\simeq1150$~K for
CO molecules, $\Delta a=2\times10^{-6}$~cm for the mantle thickness, and $p_0\simeq10^{12}$~dyne~cm$^{-2}$ for the saturated
CO-vapor pressure \citep{Leger1985}, we get $P_{\rm heat}/P_{\rm cool}\sim10^{-4}$ (for the reaction between CO and OH).

We see that the global grain cooling is much more efficient than the heating. The temperature could only increase with time
if the sublimation enthalpy $\Delta H_{\rm sub}$ would be substantially larger than the activation energy $E_{\rm a}$, say
by several hundreds of K. However, as was pointed out in Sec.~\ref{properties}, these two values are estimated to be about
the same,\footnote{Recent studies \citep{Theule2015} suggest that the diffusion energy of CO molecules in the bulk ice
(which determines the magnitude of $E_{\rm a}$) might significantly {\it exceed} the value of $\Delta H_{\rm sub}$. If so,
the ratio $P_{\rm heat}/P_{\rm cool}$ would be even smaller than estimated above.} and therefore it is rather unlikely that
the whole-grain heating could trigger the thermal explosion.

Under-critical energetic particles colliding with a grain may nevertheless stimulate reactions between radicals stored in
the mantle \citep[e.g.,][]{Reboussin2014}. To estimate this effect (assuming the whole-grain heating), we compare the
characteristic time scales of the sublimation cooling and the chemical reactions. The time to burn the characteristic
fraction $\bar{\varphi}= \sqrt{\varphi_{\rm A}\varphi_{\rm B}}$ of the major reactive species at a given temperature is
$t_{\rm chem}\sim (\bar{\varphi}\nu)^{-1} e^{E_{\rm a}/k_{\rm B}T}$, while the cooling time is $t_{\rm cool}\sim E_{\rm
dep}/P_{\rm cool}$, where $E_{\rm dep}$ is the total energy deposited in a grain. By substituting $E_{\rm dep}\simeq
2q_2a<10^4$~eV for CR protons and using otherwise the same parameters as above (and also taking into account the crossover
to the radiative cooling at lower temperatures), we obtain $t_{\rm cool}/t_{\rm chem}<3\times10^{-5}$.

We conclude that the chemical reactions stimulated by the whole-grain heating, due to collisions with under-critical
particles, are many orders of magnitude slower than the cooling. Even though the collisions with (under-critical) CR protons
are $\sim\phi_{\rm Fe}^{-1}\sim10^4$ more frequent than with (over-critical) iron ions, such reactions are not expected to
noticeably affect the chemical composition of the ice mantle at a timescale of the explosive disruption \cite[although the
abundances of trace species, such as complex organic molecules, may be changed, e.g.,][]{Reboussin2014}. It must be
stressed, however, that the above estimates completely neglect effects of the local thermal spikes generated in the mantle
by CR protons, at a timescale of thermal diffusion. The question of whether the integral effect of the heterogeneous
chemistry stimulated by such heating is more profound than that due to whole-grain heating requires a separate careful
study.

\section{Conclusions}

The main result of this article is that we have identified the regime of {\it localized} spot heating of a reactive medium,
and developed a rigorous theory describing the thermal evolution in this case. The problem is characterized by a single
dimensionless number $\lambda$ which depends on the deposited energy and properties of the medium. The theory allows us to
determine the universal explosion threshold and accurately describe impulsive heating of icy mantles by energetic particles.

A collision with an over-critical energetic particle (when $\lambda$ exceeds the threshold) leads to the thermal explosion
which, in turn, generates the flame front propagating in the mantle and leading to its disruption. We showed that heavy CR
species, such as iron ions, are able to trigger the explosion, while the stopping power of the most abundant CR protons is
insufficient for that (since the stopping power is roughly proportional to the squared atomic number of CR ions). Also, we
practically ruled out other energetic species, e.g., X rays, as possible causes of explosion due to impulsive heating.

It is important to stress that the question of how exactly the disruption occurs -- whether the mantle is completely
evaporated due to thermal explosion, or a part of it is ejected off the grain in a form of tiny ice pieces -- remains
unclear. Thus, the possibility of partial mechanical disruption of the mantle leads to a conclusion that interstellar medium
may contain solid nanoparticles of predominantly water ice.

Interestingly, the existence of a well-defined explosion threshold allows us to estimate also the upper limit of the
abundance of radicals that may be stored in the mantle: For the assumed dust and CR properties, the product of the
fractional abundances of two major radicals cannot exceed the value of $\sim3\times10^{-3}$, to avoid unrealistically large
desorption rates. Thus, the presented theory enables us to put constraints on astrochemical models.

When $\lambda$ is below the threshold, the deposited heat is quickly redistributed over the entire grain volume, i.e., the
whole-grain heating scenario is realized. The rates of reactions between radicals frozen in the mantle exponentially depend
on the temperature, so even a slight temperature increase can dramatically accelerate the release of chemical energy in the
reactive volume -- for this reason, the whole-grain heating has been considered so far as the prime possible cause of
thermal explosion. However, we have demonstrated that the explosion is unlikely in this case, since the cooling from the
grain surface (due to sublimation of volatile species) turns out to be very efficient.

The {\it non-explosive} chemical processes, induced in the mantle by under-critical impulsive heating, represent another
very important phenomenon which needs to be further investigated. We considered the whole-grain heating due to CR protons,
and demonstrated that in this case the abundance of the major reactive species is not expected to noticeably change at a
timescale of the explosive disruption (caused by heavy CR species). However, chemical reactions depend on the {\it local}
temperature and therefore evolve much faster during short transient heating events, within small volumes where
under-critical particles deposit their energy. Careful analysis of such heterogeneous chemistry (as opposed to the chemistry
due to whole-grain heating), and the evaluation of its integral effect will be reported in a future paper.

\appendix

\section{Appendix A\\ Relation between unsteady and steady problems}
\label{App_relation}

Consider a reactive medium which has a characteristic size $r_0$ and temperature $T_0$ at the boundary, and assume that
there is a thermal equilibrium. The stability of such steady state is determined by the Frank-Kamenetskii number
\citep{FrankKamenetskii,LandauFluid},
\begin{equation}\label{lambda_FK}
\lambda_{\rm FK}=\frac{Q_{\rm r}E_{\rm a}r_0^2e^{-E_{\rm a}/k_{\rm B}T_0}}{\kappa k_{\rm B}T_0^2},
\end{equation}
which is the ratio of the time scales of thermal diffusion to chemical reaction: When $\lambda_{\rm FK}$ exceeds a certain
threshold, the diffusive loss cannot compensate for temperature increase due to ongoing reaction and the steady state
becomes unstable, i.e., the thermal explosion is triggered. The thresholds for $D=1,2,$ and 3 are $\lambda_{\rm
FK,cr}=0.88,2,$ and 3.32, respectively \citep{FrankKamenetskii}.

For the unsteady problem studied in this paper, the number $\lambda$ plays a role of $\lambda_{\rm FK}$. To understand their
relation, let us calculate the ``momentary'' value of $\lambda_{\rm FK}$ for the unsteady process: The relevant scale for
$T_0$ would be the temperature $T(0,t)$ at the center of the ignition spot, while for $r_0^2$ one should substitute the
squared diffusion length $(q_D/\rho cT_0)^{2/D}$. Then, by employing Eq.~(\ref{lambda}) we get
\begin{eqnarray}
\lambda_{\rm FK}/\lambda\sim e^{-1/\theta_0}\theta_0^{-2(1+1/D)},\nonumber
\end{eqnarray}
where $\theta_0(t)=k_{\rm B}T(0,t)/E_{\rm a}$. We see that $\lambda_{\rm FK}/\lambda$ is the sole function of $\theta_0$
and, thus, of $t$. It attains maximum at $\theta_0=2(1+1/D)$, where $\lambda_{\rm FK}/\lambda\sim1$, which identifies the
``optimum moment'' to trigger the explosion (provided $\lambda>\lambda_{\rm cr}$). Physically, the optimum comes out because
the size of the reactive zone is too small in the beginning (i.e., the time scale of thermal diffusion is short), while at
later times the reaction becomes exponentially slow.

The derived relation allows us to obtain the dependence of $\lambda_{\rm cr}$ on the size of the ignition spot $w$. Using
the relation $\theta_0w^D\sim1$ (where $w$ is in units of $r_*$), we get the scaling $\lambda_{\rm cr}(w)\sim \exp(w^D)
w^{-2(1+D)}$, which provides excellent fit to the curves in Fig.~\ref{fig2} at $w\geq2$.

Thus, unlike the case of localized ignition spot, the unsteady problem for $w\gtrsim1$ is no longer characterized by a
single dimensionless number. A similar problem of thermal explosion of large ``hot spots'' has been studied numerically in
the 1960's by \cite{Merzhanov1963} and \cite{Merzhanov1966} who showed that, for a given initial size $r_0$ and temperature
$T_0$ of the spot, the explosion threshold $\lambda_{\rm FK,cr}$ has a logarithmic dependence on $T_0$.

\section{Appendix B\\ Flame front}
\label{App_front}

The explosion generates the flame front propagating away from the ignition spot. At sufficiently large times, when the front
coordinate $\xi$ is much larger than the front thickness, the last term on the rhs of Eq.~(\ref{heat_eq_norm}) becomes
asymptotically negligible (i.e., the front curvature is no longer important). Then, by separating the reactive
($\theta>\theta_{\rm tr}$) and inert ($\theta<\theta_{\rm tr}$) zones of the front \citep{FrankKamenetskii,LandauFluid}, we
can approximately describe the temperature profile by the following equation:
\begin{eqnarray}
&\theta>\theta_{\rm tr}:&\quad\frac{\partial \theta}{\partial \tau}= \lambda +\frac{\partial^2\theta}{\partial \xi^2},
\nonumber\\
&\theta<\theta_{\rm tr}:&\quad\frac{\partial \theta}{\partial \tau}= \frac{\partial^2\theta}{\partial \xi^2},\nonumber
\end{eqnarray}
where $\theta_{\rm tr}\sim1$ is the fitting parameter [to be determined from numerical solution of Eqs.~(\ref{heat_eq_norm})
and (\ref{IC_norm})]. We search the solution in the form $\theta(\xi,\tau)=\theta(s)$ with $s=\xi-u\tau$, which yields
$\theta(s)=A_1e^{-us}-(\lambda/u)s+A_2$ for $\theta>\theta_{\rm tr}$ and $\theta(s)=A_3e^{-us}$ for $\theta<\theta_{\rm
tr}$. By setting $\theta(0)=\theta_{\rm tr}$ and taking into account that in the reactive zone $\theta(s)$ cannot grow
faster than linearly, we obtain $A_1=0$; constants $A_2$ and $A_3$ are determined from continuity of $\theta$ and
$\partial\theta/\partial\xi$ at $s=0$. We get $u=\sqrt{\lambda/\theta_{\rm tr}}$ and
\begin{eqnarray}
&s<0:&\quad\theta(s)=-\sqrt{\lambda\theta_{\rm tr}}\:s+\theta_{\rm tr},\nonumber\\
&s>0:&\quad\theta(s)=\theta_{\rm tr}\exp\left(-\sqrt{\lambda/\theta_{\rm tr}}\:s\right),\nonumber
\end{eqnarray}
the numerical fit yields $\theta_{\rm tr}\simeq1.3$. In physical units, the front speed,
\begin{equation}\label{U}
U=\sqrt{\frac{Q_{\rm r}\kappa}{\theta_{\rm tr}(\rho c)^2E_{\rm a}}},
\end{equation}
is determined by the reactive and transport properties of the medium.

\section{Appendix C\\ Thermal explosion when $c$ or $\chi$ are functions of $T$}
\label{App_T_dependence}

Let us consider the case when the specific heat is a function of temperature, $c(T)$, while the thermal diffusivity
$\chi=\kappa/\rho c$ is first assumed to be constant. It is convenient \citep{LandauFluid} to introduce the enthalpy
$H=\rho\int c\:dT\equiv F(T)$, noting that $F(T)$ is a single-valued (monotonously increasing) function. Then
Eqs.~(\ref{heat_eq}) and (\ref{IC}) can be rewritten in the following identical form for $H$:
\begin{eqnarray}
\frac{\partial H}{\partial t}= Q_{\rm r}e^{-E_{\rm a}/k_{\rm B}T}+\chi\left(\frac{\partial^2H}{\partial r^2}+\frac{D-1}r
\frac{\partial H}{\partial r}\right),\nonumber\\
H(r,0)=q_D\delta_D(r),\nonumber
\end{eqnarray}
where $T=F^{-1}(H)$ is the inverse function. We introduce the enthalpy scale, $H_E=F(E_{\rm a})$, and conclude that the
problem can be reduced to the dimensionless form of Eqs.~(\ref{heat_eq_norm}) and (\ref{IC_norm}), where (apart from the
Arrhenius term) $H/H_E$ should be substituted for $\theta$; in the Arrhenius term, $\theta$ should be replaced with
$F^{-1}(H/H_E)$, and
\begin{equation}\label{lambda1}
\lambda=\frac{Q_{\rm r}}{\chi H_E}\left(\frac{q_D}{H_E}\right)^{2/D}.
\end{equation}
Correspondingly, the speed of the flame front is given by
\begin{equation}\label{U1}
U=\sqrt{\frac{Q_{\rm r}\chi}{\theta_{\rm tr}H_E}}.
\end{equation}
For example, for $c\propto T^{\alpha}$ with the exponent $\alpha\geq0$, we get $H/H_E=\theta^{1+\alpha}$, where $H_E=\rho
c(E_{\rm a})E_{\rm a}/(1+\alpha)$; for a constant specific heat ($\alpha=0$) Eqs.~(\ref{lambda1}) and (\ref{U1}) are reduced
to Eqs.~(\ref{lambda}) and (\ref{U}), respectively. Figure~\ref{fig3} shows that the explosion threshold decreases
dramatically with $\alpha$.

Also, we analyzed the effect of the temperature-dependent thermal diffusivity $\chi$. In this case the diffusion term in the
heat equation becomes nonlinear. From the numerical solution with $\chi\propto T^{\beta}$ we obtained the dependencies
$\lambda_{\rm cr}(-\beta)$ that are qualitatively similar to $\lambda_{\rm cr}(\alpha)$ shown in Fig.~\ref{fig3} (i.e.,
$\lambda_{\rm cr}$ monotonically increases with $\beta$). However, the relative variation of $\lambda_{\rm cr}$ with $\beta$
turns out to be several times smaller than with $\alpha$, i.e., the effect of $\chi(T)$ on the explosion threshold is
substantially weaker than that of $c(T)$.

\bibliographystyle{apj}
\bibliography{refs}

\end{document}